\documentclass[preprintnumbers, showpacs, floatfix,onecolumn,preprintnumbers, letterpaper, superscriptaddress,nofootinbib]{revtex4}
\usepackage[utf8]{inputenc}
\usepackage{amsfonts}
\usepackage{amsmath}
\usepackage{amssymb,epsf}
\usepackage{latexsym}
\usepackage{graphicx,epsfig}
\usepackage{amssymb}
\usepackage{float}
\usepackage{subfigure}
\usepackage{epstopdf}
\usepackage[colorlinks=true,citecolor=blue,linkcolor=blue,urlcolor=black]{hyperref}
\usepackage{dcolumn}
\usepackage{psfrag}
\usepackage{wrapfig}
\usepackage{makeidx}
\usepackage{epsf}

\begin{document}

    \title{Holographic paramagnetic-ferromagnetic phase transition of Power-Maxwell-Gauss-Bonnet black holes}
    \author{B. Binaei Ghotbabadi}
    \affiliation{Physics Department, College of Sciences, Shiraz
        University, Shiraz 71454, Iran}

    \affiliation{Biruni Observatory, College of Sciences, Shiraz
        University, Shiraz 71454, Iran}

    \author{A. Sheykhi}
    \email{asheykhi@shirazu.ac.ir}

    \affiliation{Physics Department, College of Sciences, Shiraz
        University, Shiraz 71454, Iran}

     \affiliation{Biruni Observatory, College of Sciences, Shiraz
        University, Shiraz 71454, Iran}

    \author{G. H. Bordbar} \email{ghbordbar@shirazu.ac.ir}
    \affiliation{Physics Department, College of Sciences, Shiraz
        University, Shiraz 71454, Iran}

       \author{A. Montakhab}
     \affiliation{Physics Department, College of Sciences, Shiraz
        University, Shiraz 71454, Iran}

    \begin{abstract}
         Based on the shooting method, we numerically investigate the
        properties of holographic paramagnetism-ferromagnetism phase
        transition in the presence of higher order Gauss-Bonnet (\emph{GB})
        correction terms on the gravity side. On the matter field side, however, we
        consider the effects of the Power-Maxwell (\emph{PM}) nonlinear
        electrodynamics on the phase transition of this system. For this
        purpose, we introduce a massive $2-$form coupled to \emph{PM} field, and
        neglect the effects of $2-$form fields and gauge field on the
        background geometry. We observe that increasing the strength of
        both the power parameter $q$ and \emph{GB} coupling constant $\alpha$
        decrease the critical temperature of the holographic model, and
        lead to the harder formation of magnetic moment in the black hole
        background. Interestingly, we find out that at low
        temperatures, the spontaneous magnetization and ferromagnetic
        phase transition happen in the absence of external magnetic field.
        In this case, the critical exponent for magnetic moment has
        the mean field value, $1/2$, regardless of the values of $q$ and
        $\alpha$. In the presence of external magnetic field, however, the
        magnetic susceptibility satisfies the Curie-Weiss law.
    \end{abstract}

    \maketitle

    \section{Introduction}
    The duality between gravity in an anti-de Siter (AdS) spacetime
    and a conformal field theory (CFT), known as AdS/CFT
    correspondence, provides a powerful tool to investigate strongly
    coupled systems \cite{Maldacena,Gubser,HorowitzM,Witten,Ren}. A
    significant applications of this conjecture is investigation of
    the electronic properties of materials and magnetism
    \cite{hartnoll,Herzog,McGreevy,HerzogA,GubserB,montull,Donos,albash,m.pujo,iqbal}.
    The gauge/gravity duality, which is a new approach for calculating
    the properties of superconductors using a dual classical gravity,
    also provides a fascinating tool to shed light on high
    temperature superconductors. Based on this theory, one can
    describe a superconductor using a dual classical gravity
    description. It has been shown that some properties of strongly
    coupled superconductors can be potentially described by classical
    general relativity living in one higher dimension, which is known
    as holographic superconductors \cite{hartnolBuild}. The idea of
    holographic superconductor was initiated by Hartnol, et. al.
    \cite{hartnolBuild, hartnolSup}. They considered a
    four-dimensional Schwarzschild-AdS black hole coupled to a Maxwell
    and a scalar fields to construct a holographic $s$-wave
    superconductor. Based on their model, in order to describe a
    holographic superconductor on the boundary, a transition from
    hairy black hole to a no hair black hole in the bulk for
    temperatures below and upper the critical value is required
    \cite{hartnolBuild}. The appearance of hair corresponds to the
    spontaneous $U(1)$ symmetry breaking \cite{hartnolBuild}. This
    theory opened up a new perspective in condensed matter physics to
    study the high temperature superconductors. During the past
    decade, the explorations on the holographic dual models have
    attracted considerable attention (see e.g. \cite{Lai,
        Sh1,Sh2,Sh3,Rogatko,Kuang,Mansoori,Ling,Introduction Cai,Lifshitz
        Wu,Bahareh,Mahya1,Mahya2,Mahya3,Mahya4,dehyadgari} and reference therein).
    The studies were also generalized to investigate
    paramagnetic-ferromagnetic phase transition using the holographic
    description
    \cite{dyonic,p.Acai6,Coexistence.Cai,Yokoi,Cai3,Cai4,Insulator.Cai,Understanding.Cai,Lifshitz5,Zhang2,Wu1,binaei1}.
    The first holographic paramagnetic-ferromagnetic phase transition
    model was a dyonic Reissner-Nordstrom-AdS black brane
    \cite{dyonic}. This model provides a starting point for
    exploration of more complicated magnetic phenomena and quantum
    phase transition, by considering a real antisymmetric tensor field
    which is coupled to the background gauge field strength in the
    bulk. It was argued that the spontaneous magnetization which
    happens in the absence of an external magnetic field, can be
    realized as the paramagnetic-ferromagnetic phase transition. Most
    investigations on holographic paramagnetism-ferromagnetism phase
    transition have been carried out by considering the gauge field as
    a linear Maxwell field in Einstein gravity
    \cite{p.Acai6,Coexistence.Cai,Yokoi,Cai3,Cai4,Insulator.Cai,Understanding.Cai,Lifshitz5}.
    It is also of great interest to explore the effects of nonlinear
    electrodynamics on the properties of the holographic
    ferromagnetic-paramagnetic phase transition. These holographic
    setups have been widely studied in the presence of nonlinear
    electrodynamics, those involve more information than the usual
    Maxwell state \cite{Zhang2,Wu1}. It has been observed that in the
    Schwarzschild AdS black hole background and in the absence of
    external magnetic field, the higher nonlinear electrodynamics
    corrections make the magnetization harder to be formed. Among
    various nonlinear extension of Maxwell electrodynamics, the PM
    nonlinear electrodynamics which preserves the conformally
    invariant feature in higher dimensions has received more
    attraction \cite{shamsip30}. The conformally invariant PM action
    in $(n+1)$-dimensional may be written,
    \begin{equation} \label{act2}
    I_{PM}=\int{d^{n+1}x \sqrt{-g}\left(-\mathcal{F}\right)^q},
    \end{equation}
    where $\mathcal{F}=F_{\mu\nu}F^{\mu\nu}$ is the Maxwell invariant
    and $q$ is the power parameter. Under conformal transformation
    $g_{\mu\nu}\to\Omega^{2}g_{\mu\nu}$ and $A_{\mu}\to A_{\mu}$, the
    above action remains invariant. The corresponding energy-momentum
    tensor given by
    \begin{equation}
    T_{\mu\nu}=2\left(q
    F_{\mu\rho}F{^{\rho}_{\nu}}\mathcal{F}^{q-1}-\frac{1}{4}g_{\mu\nu}\mathcal{F}^{q}\right),
    \end{equation}
    is traceless for $4q=n+1$. Black hole solutions in the presence of
    PM electrodynamics have been constructed by many authors (see e.g.
    \cite{Hendi1,SheyPM1,SheyPM2,SheyPM3,kordzangeneh1,kordzangeneh2,
        kordzangeneh3} and references therein). The
    properties of holographic superconductor with conformally
    invariant $PM$ electrodynamics have been studied in
    Refs.\cite{PM1,PM2,Shey1,Shey2,Shey3,Shey4,Doa}. Recently, we
    explored the effects of \emph{PM} nonlinear electrodynamics on the
    properties of holographic paramagnetic-ferromagnetic phase
    transition in the background of Schwarzchild-AdS black hole
    \cite{binaei1}. We investigated how the \emph{PM} electrodynamics
    influences the critical temperature and magnetic moment. We found
    that the effects of $PM$ field lead to the easier formation of
    magnetic moment at higher critical temperature. The studies were
    also generalized to other gravity theories. In the context of $GB$
    gravity, the phase transition of the holographic superconductors
    were explored in Refs.\cite
    {q.pan,Nie,CaiGauss,JingG,Pan.Analytical,Pan.Holographic,Pan.General,Li.Gauss,Barclay}.
    Their motivations are to study the effects of higher order gravity
    corrections on the critical temperature of holographic
    superconductors. They found that when $GB$ coefficients become
    larger, the condensation on the boundary field theory
        becomes harder to be formed. In our previous work
    \cite{binaei1} we have considered the effects of $PM$ on
    paramagnetic-ferromagnetic phase transition in Einstein gravity.
    It it also interesting to examine the effects of this kind of
    nonlinear electrodynamics when the higher order corrections on the
    gravity side such as GB terms, is taken into account. We would
    like to examine whether or not the holographic
    paramagnetic-ferromagnetic phase transition still hold in the
    presence of higher order gravity corrections. We shall apply the
    shooting method to numerically investigate the influences of both the
    higher order $GB$ curvature correction terms, as well as the
    nonlinear $PM$ electrodynamics on the holographic system.

    This paper is organized as follows. In section \ref{setup}, we
    introduce the action and basic field equations in the presence of
    PM electrodynamics by considering the higher order $GB$ curvature
    correction terms. In section \ref{numst}, we employ the shooting
    method to obtain numerically critical temperature and magnetic
    moment of the system. We investigate the effect of external magnetic field for our model and obtain the magnetic susceptibility
    density in section \ref{suscept}. In the last section, we finish with closing remarks.
    \section{Holographic Model\label{setup}}
    In this section, we introduce the action of Einstein-Gauss-Bonnet in AdS spaces which is coupled to a $PM$ field. The action of an $(n+1)-$dimensional Einstein-Gauss-Bonnet gravity can be written as,
    \begin{eqnarray}
    S&=&\frac{1}{16\pi G}\int
    d^{n+1}x\sqrt{-g}\left[R-{2}{\Lambda}+\frac{\alpha}{2}
    (R^{2}-4R^{\mu\nu}R_{\mu\nu}+R^{\mu\nu\rho\sigma}R_{\mu\nu\rho\sigma})
    + L_{1}\left(
    \mathcal{F}\right)
    +\lambda^{2} L_{2}\right],  \notag \\
    &&\label{Act}
    \end{eqnarray}%
    with
    \begin{eqnarray}{\nonumber}
    L_{2}=-\frac{1}{12}(dM)^{2}-\frac{m^{2}}{4}M_{\mu\nu}M^{\mu\nu}-%
    \frac{1}{2}M^{\mu\nu}F_{\mu\nu}-\frac{J}{8}V(M),
    \label{Act1}
    \end{eqnarray}
    where $G$ is Newtonian gravitational
    constant, $g$ is the determinant of metric and $\alpha$ is the
    $GB$ coefficient. In the above action, $R, R_{\mu\nu},
    R_{\mu\nu\rho\sigma}$ are, respectively, Ricci scalar, Ricci
    tensor and Riemann curvature tensor. Here $\lambda$ and $J$ are two real  parameters. $\lambda^{2}$ characterizes the
    back reaction of the $2-$form field $M_{\mu\nu}$ to the
    Maxwell field strength and to the background geometry. In addition,
    $m$ is the mass of the real tensor field $M_{\mu\nu}$, being greater than zero \cite{Cai3}, and $dM$ is the exterior differential $2$-form field $M_{\mu\nu}$.
     In the above action, when $\alpha\to0$, the Enistein-Maxwell
    theory is recovered \cite{binaei1}. In addition
    $\Lambda=-{n(n-1)}/{2l^{2}}$ is the cosmological constant of
    $(n+1)$-dimensional AdS spacetime with radius $l$. Here
    $L_{1}(\mathcal{F})=-\mathcal{F}^{q}/4$, where $\mathcal{F}=F_{\mu
        \nu }F^{\mu \nu }$ in which $F_{\mu \nu }=\partial_{\mu} A_{\nu }-\partial_{\nu} A_{\mu }$ and $A_{\mu}$ is the Maxwell potential and $q$ is the power parameter of the $PM$ field. In the
    limiting case, $q\to 1$, the $PM$ Lagrangian will reduce to the
    Maxwell case, $L_{1}=-F_{\mu\nu}F^{\mu\nu}/4$. The theory is conformally invariant for $q={(n+1)}/{4}$ and the energy-momentum tensor of the $PM$ Lagrangian
    is traceless in all dimension\cite{shamsip30}.

     $V(M_{\mu\nu})$ describes the self interaction of polarization tensor. This
    nonlinear potential should be expanded as the even power of
    $M_{\mu\nu}$. For simplicity, we take the following form for this model, \cite{Cai4}
    \begin{equation}
    V\left(M\right) =\left( ^{*}M_{\mu\nu}{M^{\mu\nu}}\right)^{2}=[^{*}(M \wedge M)]^{2},
    \label{potential}
    \end{equation}%
    where $*$ is the Hodge star operator. Since
    we consider the probe limit, the gauge and matter fields do not
    back react on the background metric. The line element of the
    metric with flat horizon is given by \cite{CaiGB}
    \begin{equation}
    ds^{2}=-r^{2}f(r)dt^{2}+\frac{dr^{2}}{r^{2}f(r)}+r^{2}%
    \sum_{i=1}^{n-1}dx_{i}^{2},
    \end{equation}%
    with
    \begin{equation}
    f(r)=\frac{1}{2\alpha}\left[1-\sqrt{1-4\alpha\left(1-\frac{r_{+}^n}
        {r^n}\right)}\right],
    \end{equation}%
    where $r_{+}$ is the positive real root of Eq. $f(r_{+})=0$. At
    large distance where $r\rightarrow \infty$, the metric function
    reduces to
    \begin{equation}
    f(r)\approx\frac{1}{2\alpha}\left[1-\sqrt{1-4\alpha}\right],
    \end{equation}%
    Since $f(r)$ should be a positive definite function, it implies
    $0\leq\alpha\leq {1}/{4}$ and real, where the upper bound
    $\alpha=1/4$ is called the Chern-Simon limit \cite{Nie}. We can
    present the effective AdS radius $L_{\rm eff}$ as
    \begin{equation}
    L^{2}_{\rm eff}=\frac{2 \alpha}{1-\sqrt{1-4\alpha} }. \label{3}
    \end{equation}%
    The Hawking temperature associated with the  black hole event
    horizon, which can be interpreted as the temperature of CFT on the
    boundary, is given by \cite{q.pan}
    \begin{equation}
    T=\frac{f^{\prime }(r_{+})}{4\pi },
    \label{3}
    \end{equation}%
    Varying action (\ref{Act}) with respect to $2-$form field,
    $M_{\mu\nu}$, and the gauge field, $A_{\nu}$, we arrive at the
    following field equations \cite{binaei1}
    \begin{eqnarray}
    0&=& \nabla ^{\tau }(dM)_{\tau\mu\nu}-m^{2}M_{\mu\nu}-J (^{*}M_{\tau\sigma}M^{\tau\sigma})(^{*}M_{\mu\nu})%
    -F_{\mu\nu},  \label{01} \\
    0&=& \nabla ^{\mu }\left(q{F_{\mu\nu}}{(\mathcal{F})^{q-1}}+\frac{\lambda^{2}}{4}%
    M_{\mu\nu}\right).  \label{02}
    \end{eqnarray}
    Our aim here is to investigate the effects of the power parameter
    $q$ and the $GB$ coefficient $\alpha$ on the holographic
    ferromagnetic-paramagnetic phase transition. We adopt the
    following self-consistent ansatz for the matter and gauge fields \cite{Cai4}
    \begin{equation}
    M_{\mu\nu}=-p(r)dt{\wedge}dr+ \rho(r)dx{\wedge}dy ,  \label{M}
    \end{equation}%
    \begin{equation}
    A_{\mu}=\phi(r)dt+ B xdy,  \label{A}
    \end{equation}%
    where $B$ is a uniform magnetic field viewed as an
    external magnetic field in the boundary field theory. Here
    $\rho(r)$ and $p(r)$ are the components of $M_{\mu\nu}$, while
    $\phi(r)$ stands for the electric potential. Inserting the above
    ansatz into Eqs. (\ref{01}) and (\ref{02}), lead to the following nontrivial equations of motion,
    \begin{eqnarray}
    0 &=&\rho ^{\prime \prime }+\rho ^{\prime }\left[ \frac{%
        f^{\prime }}{f}+\frac{n-3}{r}\right] -\frac{\rho
    }{r^{2}f}\left[m^{2}+4Jp^{2}\right]+\frac{B}{r^{2}f},\nonumber\\
    0 &=&\left(m^{2}-\frac{4J\rho^{2}}{r^4}\right)p-\phi
    ^{\prime},\label{EOM}\\
    0&=& \phi ^{\prime \prime}+\frac{2
        \phi^{\prime}}{r}\left[\frac{\frac{n-1}{2}\phi^{\prime 2}+
        \left(2q-\frac{n+3}{2}\right)\frac{B^{2}}{r^{4}}}{(2q-1)\phi^{\prime
            2}-\frac{B^{2}}{r^{4}}}\right]+\frac{\lambda^{2}}{q 2^{q+1}
    }\left(p^{\prime}+\frac{(n-1)p}{r}\right)\left[\frac{\left(\phi^{\prime
            2}-\frac{B^{2}}{r^{4}}\right)^{2-q}}{(2q-1)\phi^{\prime
            2}-\frac{B^{2}}{r^{4}}}\right],\nonumber
    \end{eqnarray}%
    here a prime denotes the derivative with respect to $r$. Obviously, for Einstein gravity ($\alpha\rightarrow 0$) in the Maxwell limit where $q\rightarrow 1$ and $n=3$, the above equations reduce to the standard holographic ferromagnetic-paramagnetic phase transition models discussed
    in Ref.~\cite{Cai3}.
    In order to solve Eqs.(\ref{EOM}) numerically, we should specify
    the boundary conditions for the fields. Imposing the regularity
    conditions at the horizon($r=r_{+}$), yields the following
    boundary conditions \cite{dyonic}
    \begin{equation}
    \phi^{\prime}(r_{+})=(m^{2}-4J \rho^{2})p,\qquad
    \rho^{\prime}(r_{+})=\frac{\rho(r_{+})(m^{2}+4Jp^{2})-B}{4\pi T}.
    \end{equation}%
    Since the behaviors of model functions are asymptotically $AdS$,
    thus we solve the field equations (\ref{EOM}) near the boundary
    ($r\rightarrow \infty $). We find the asymptotic solutions as
    \begin{gather}
    \phi (r)\sim \mu
    -\frac{\sigma}{r^{\frac{(n-1)}{2q-1}-1}},\qquad
    p(r)\sim \frac{(\frac{n-2q}{2q-1}){\sigma}}{r^{\frac{(n-1)}{2q-1}}},\notag\\
    \rho (r)\sim \frac{\rho _{-}}{%
        r^{\Delta _{-}}}+\frac{\rho _{+}}{r^{\Delta
            _{+}}}+\frac{B}{m^{2}}, \label{boundval}
    \end{gather}%
    with
    \begin{equation}
    \Delta _{\pm }=\frac{1}{2}\left[-(n-4)\pm\sqrt{(n-4)^{2}+4m^{2}L^{2}_{\rm
            eff}}\right].
    \end{equation}
    here $\rho_{\pm}$, $\mu$ and $\sigma$ are all constants.
    Based on the AdS/CFT correspondence, $\rho_{+}$ and $\rho_{-}$ are
    interpreted as the source and vacuum expectation value of the dual
    operator. According to Ref.\cite{dyonic}, we consider $\rho_{+}$ as the source of the dual operator when $B=0$, which plays the role of order parameter in
    the boundary theory. Moreover, $\mu $ and $\sigma$ are regarded as
    the chemical potential and charge density of dual field theory,
    respectively. Unlike other nonlinear electrodynamics \cite{Zhang2}, the boundary condition for the gauge field $\phi$ depends on the power parameter $q$ of the $PM$ electrodynamic. To find the restriction to this parameter,
    we require to confined $\phi$ near boundary($r \to \infty$), so we set $\frac{n-1}{2q-1}-1>0$, which
    implies that the power parameter $q$ ranges as $1/2<q<{n}/{2}$. In
    the next section, we solve the field equations numerically and
    obtain the physical properties of our holographic model.
    \begin{figure*}
        \centering{%
            \subfigure[~$\protect$$\alpha$=0]{
                \label{fig1a}\includegraphics[width=.28\textwidth]{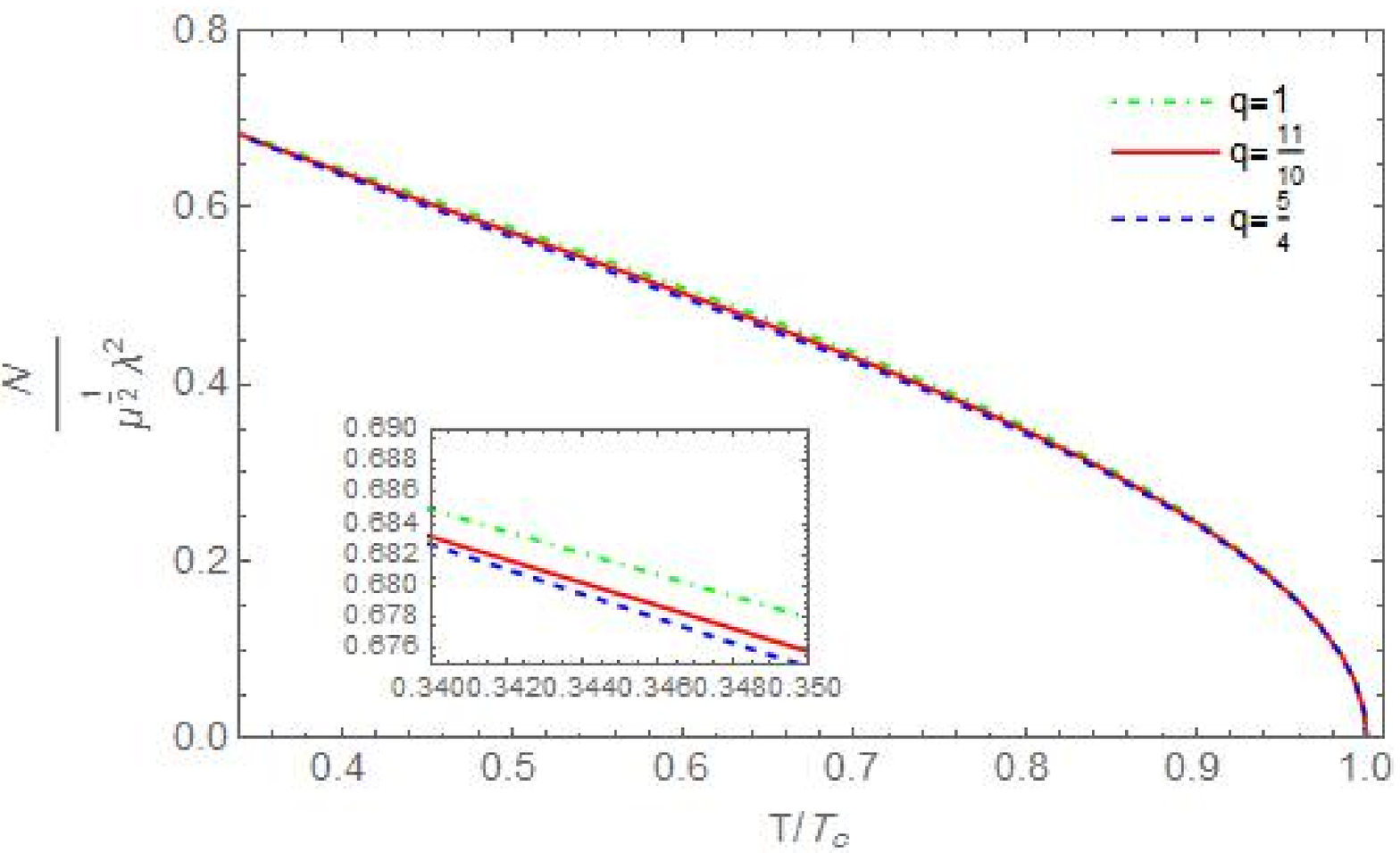}\qquad}%
            \subfigure[~$\protect$ $\alpha$=0.1]{
                \label{fig1b}\includegraphics[width=.28\textwidth]{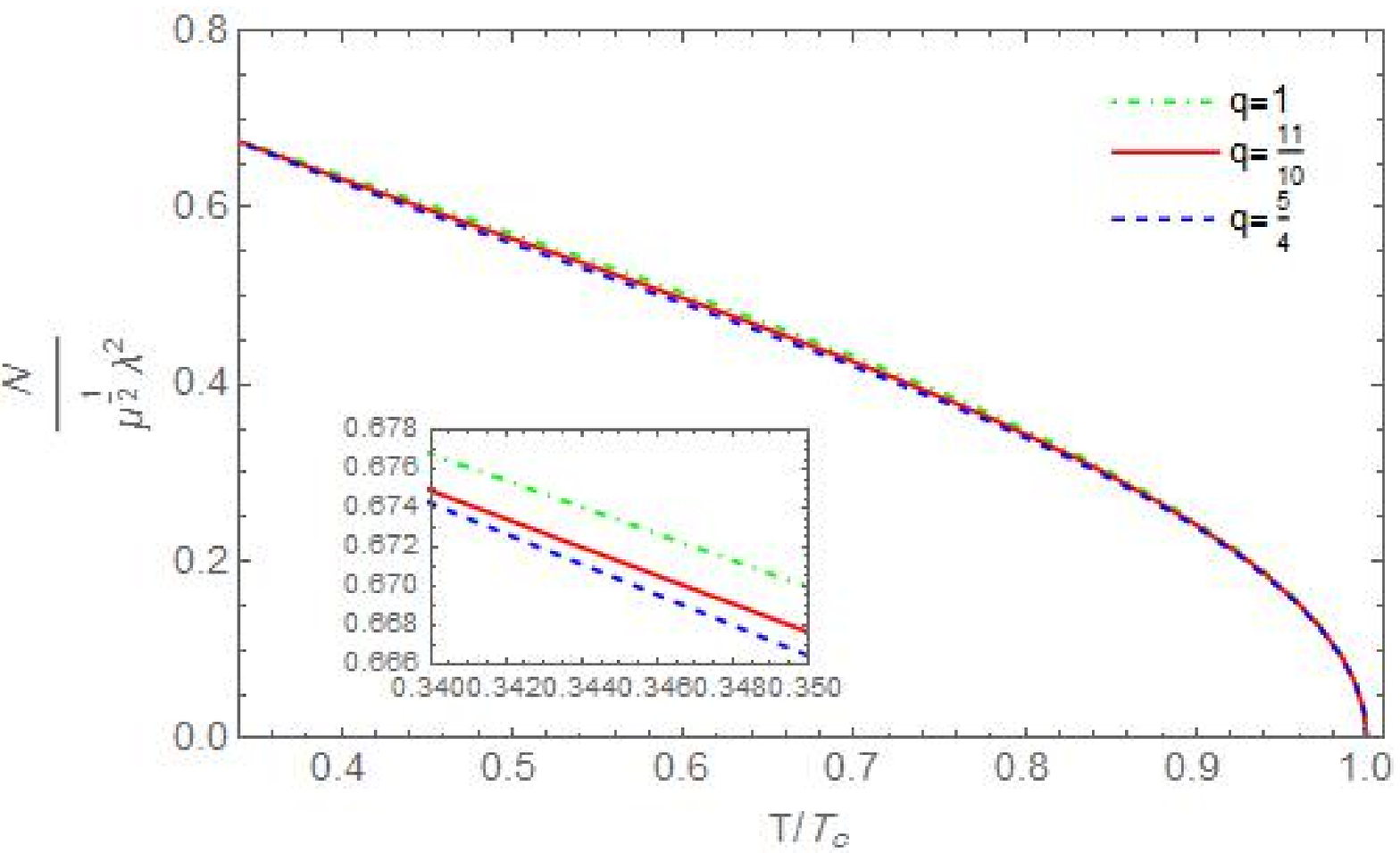}\qquad}
            \subfigure[~$\protect$$\alpha$=0.2]{
                \label{fig1c}\includegraphics[width=.28\textwidth]{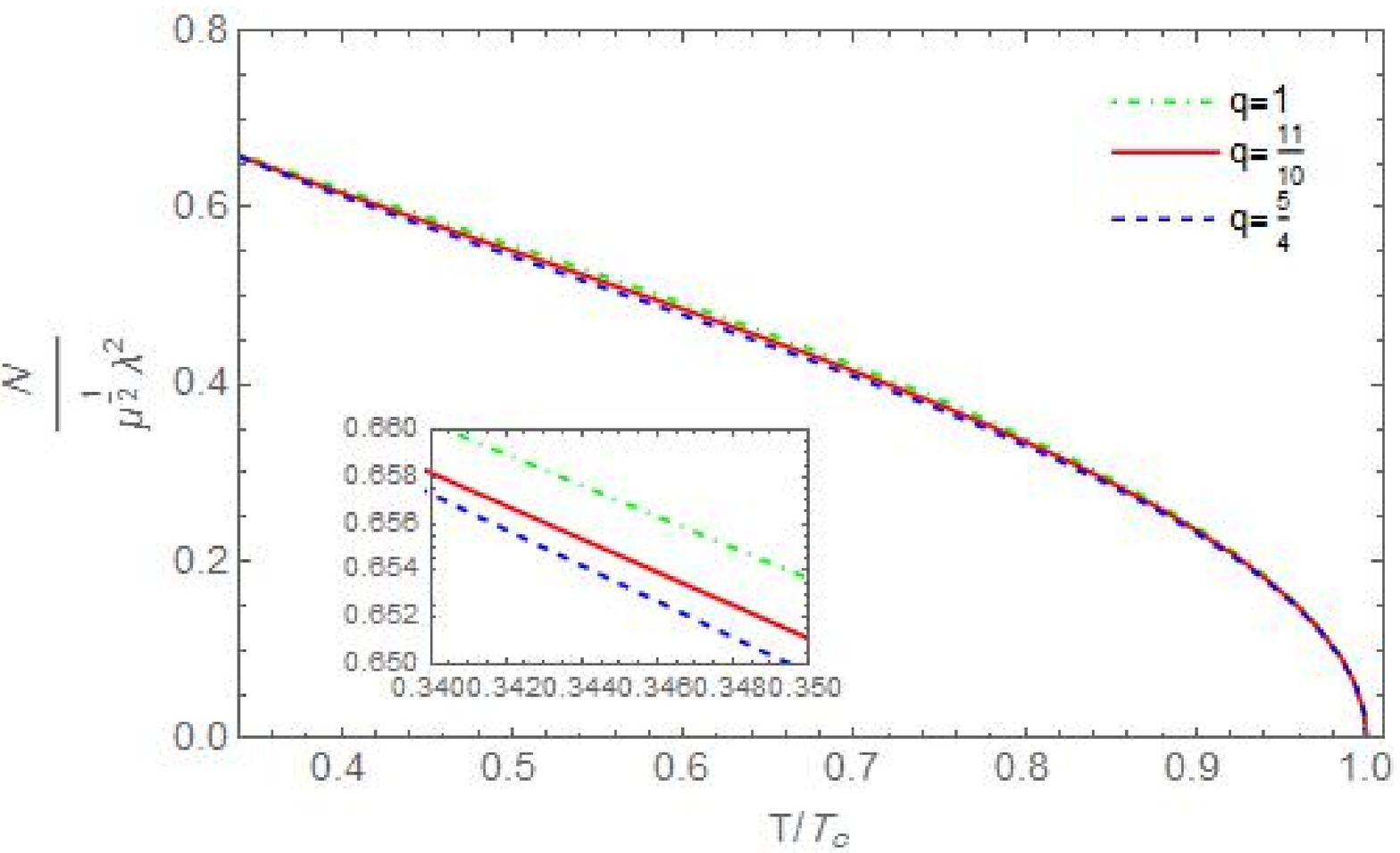}\qquad}}%

        \caption{The behavior of magnetic moment $N$ and the critical
            temperature  with different values of power parameter $q$ for $\alpha=0,0.1$ and $0.2$ in $(n+1=)5-$ dimensions. Here we have taken $m^{2}=1/8$ and $J=-1/8$.}
        \label{fig1}
    \end{figure*}

    \begin{figure*}
        \centering{%
            \subfigure[~$\protect$$q$=1]{
                \label{fig2a}\includegraphics[width=.28\textwidth]{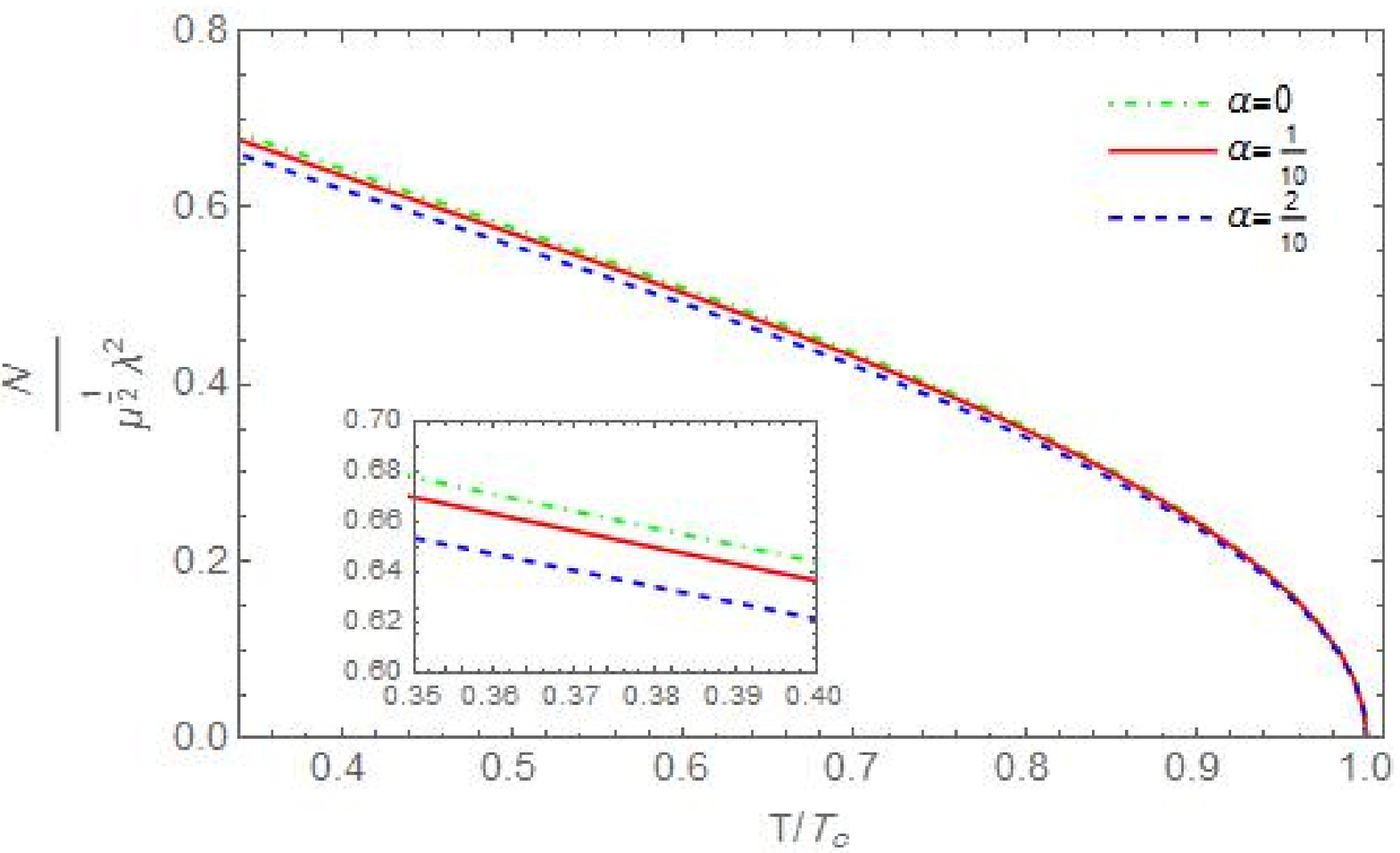}\qquad}%
            \subfigure[~$\protect$ $q$=11/10]{
                \label{fig2b}\includegraphics[width=.28\textwidth]{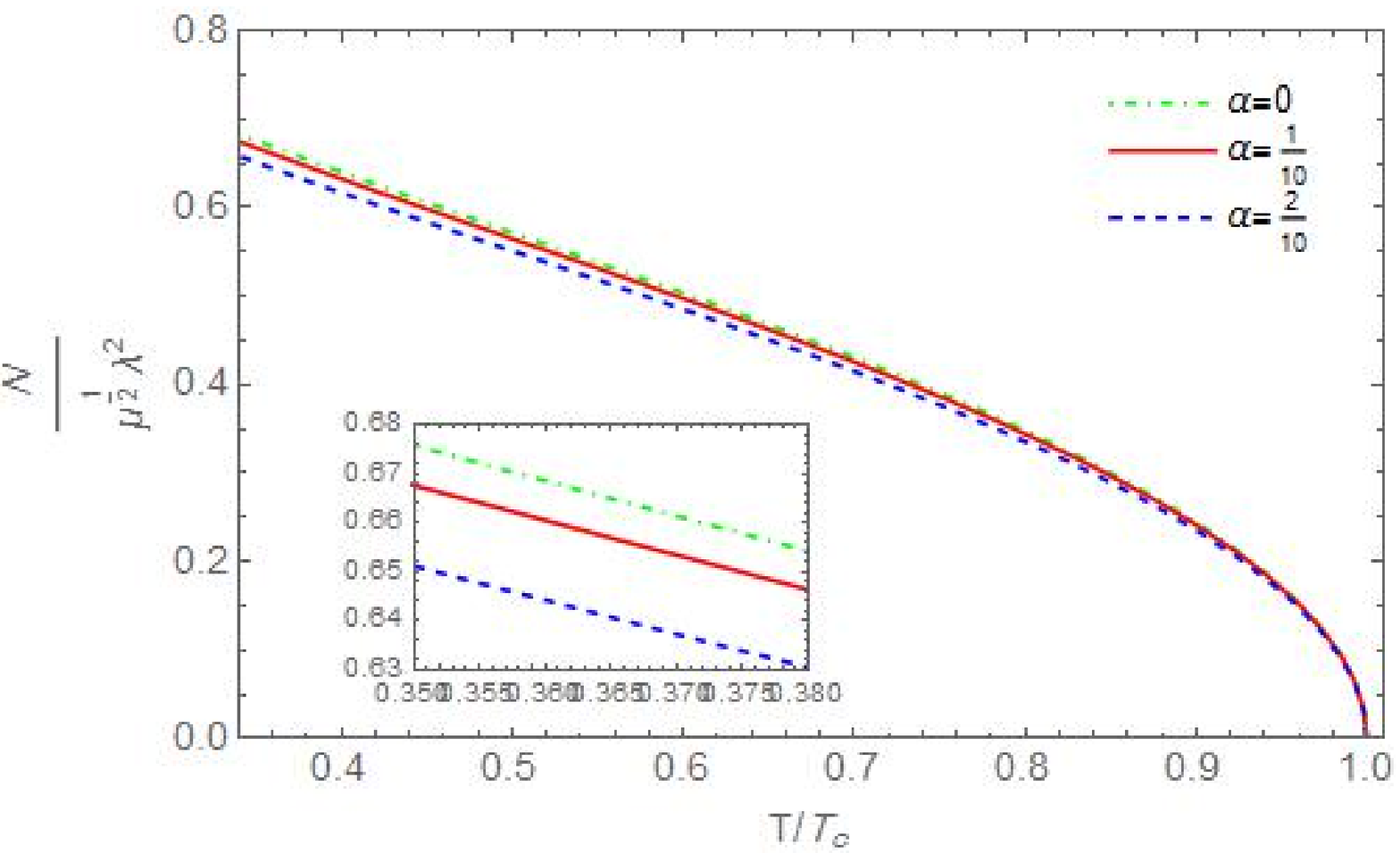}\qquad}
            \subfigure[~$\protect$$q$=5/4]{
                \label{fig2c}\includegraphics[width=.28\textwidth]{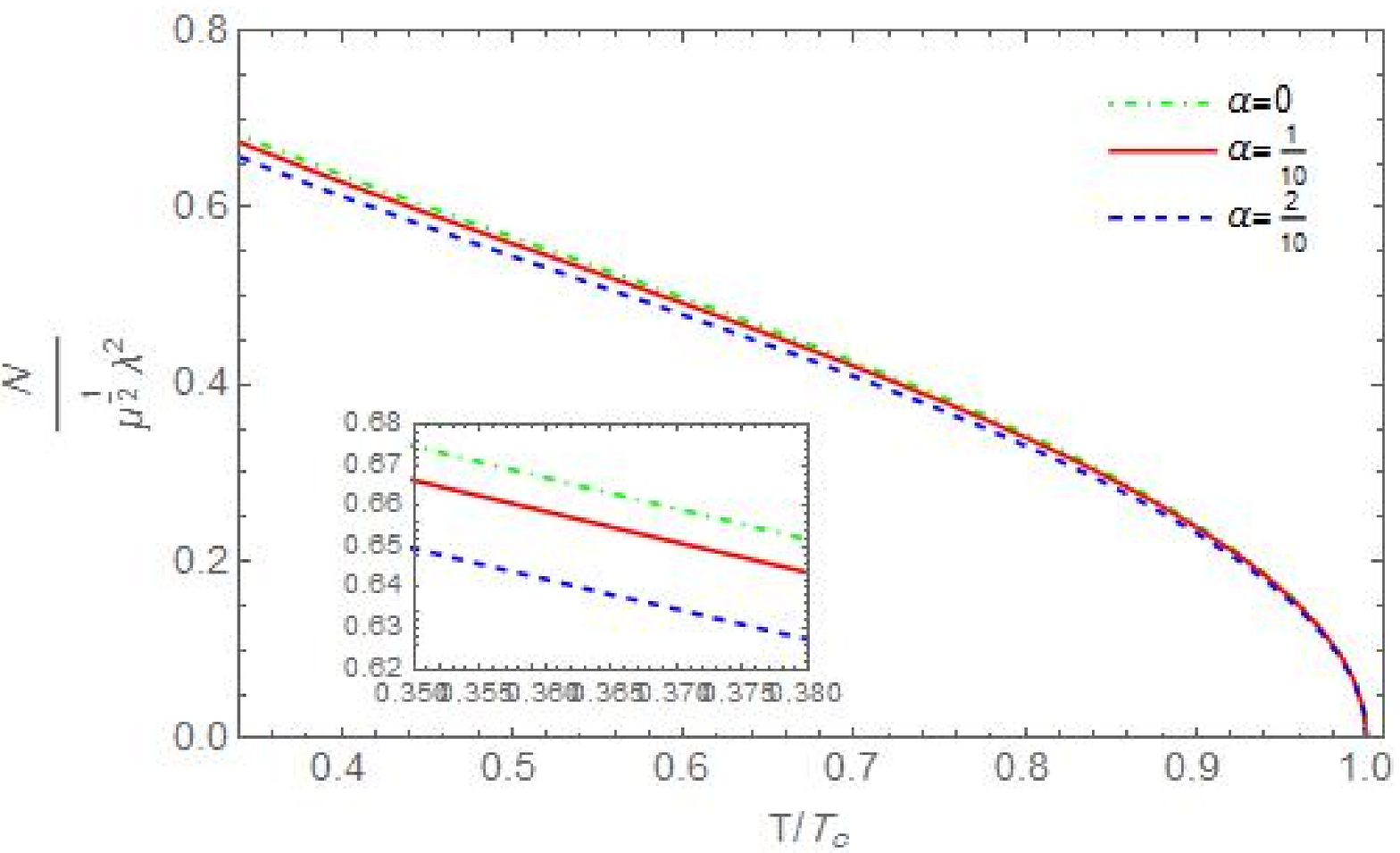}\qquad}}%

        \caption{The behavior of magnetic moment $N$ and the critical
            temperature for different values of $GB$ parameter $\alpha$ with $q=1,11/10,5/4$.
            Here we have taken $n=4$, $m^{2}=1/8$ and $J=-1/8$.}
        \label{fig2}
    \end{figure*}

    \begin{table*}[ht]
        \centering%
        \caption{Numerical results of ${T_{c}}/{\protect\mu }$ for
            different values of $q $ and $\alpha$, in $5D$.}
        \begin{tabular}{llllll}
            \hline
            $q$ &  $ 3/4 $&  $ 1 $& $ 11/10$ & $5/4 $   \\
            \hline
            $\alpha=0 $&  $ 3.3494 $ &  $2.4368 $& $2.3177$ & $2.2604$   \\
            $\alpha=0.1 $&  $ 3.1568 $ &  $ 2.3070 $& $2.1938$ & $2.1375$   \\
            $\alpha=0.2 $&  $ 2.8503 $ &  $2.0982  $& $1.9943$ & $1.9393$   \\
            \hline

        \end{tabular}

        \label{Table1}

    \end{table*}

    \begin{table*}[ht]
        \centering%
        \caption{The magnetic moment $N$ with different values of power parameter $q$
            and $GB$ coefficient $\alpha$ in $5D$.}
        \begin{tabular}{llllll}
            \hline
            $q$ &  $1 $& $ 11/10$ & $5/4 $   \\
            \hline
            $\alpha=0 $ &  $ 1.7084(1-T/Tc)^{1/2} $& $1.7095(1-T/Tc)^{1/2}$ & $1.7174(1-T/Tc)^{1/2}$   \\
            $\alpha=0.1 $ &  $ 1.6879(1-T/Tc)^{1/2} $& $1.6887(1-T/Tc)^{1/2}$ & $1.6960(1-T/Tc)^{1/2}$   \\
            $\alpha=0.2 $ &  $ 1.6466(1-T/Tc)^{1/2} $& $1.6467(1-T/Tc)^{1/2}$ & $1.6531(1-T/Tc)^{1/2}$   \\
            \hline

        \end{tabular}

        \label{Table2}

    \end{table*}

    \begin{figure*}[ht]
        \centering{%
            \subfigure[~$\protect$$q=9/10,\alpha=$0]{
                \label{fig32a}\includegraphics[width=.26\textwidth]{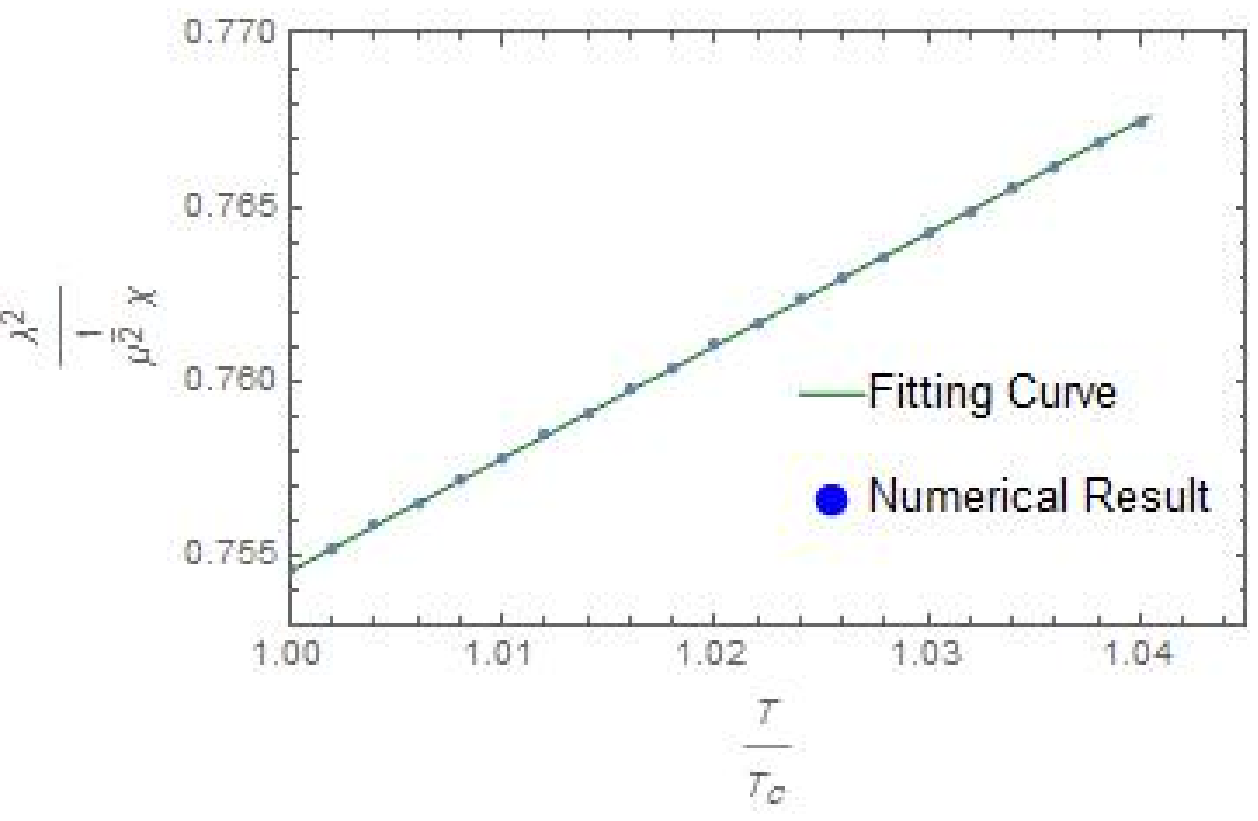}\qquad}
            \subfigure[~$\protect$$q=1,\alpha=$0]{
                \label{fig32b}\includegraphics[width=.26\textwidth]{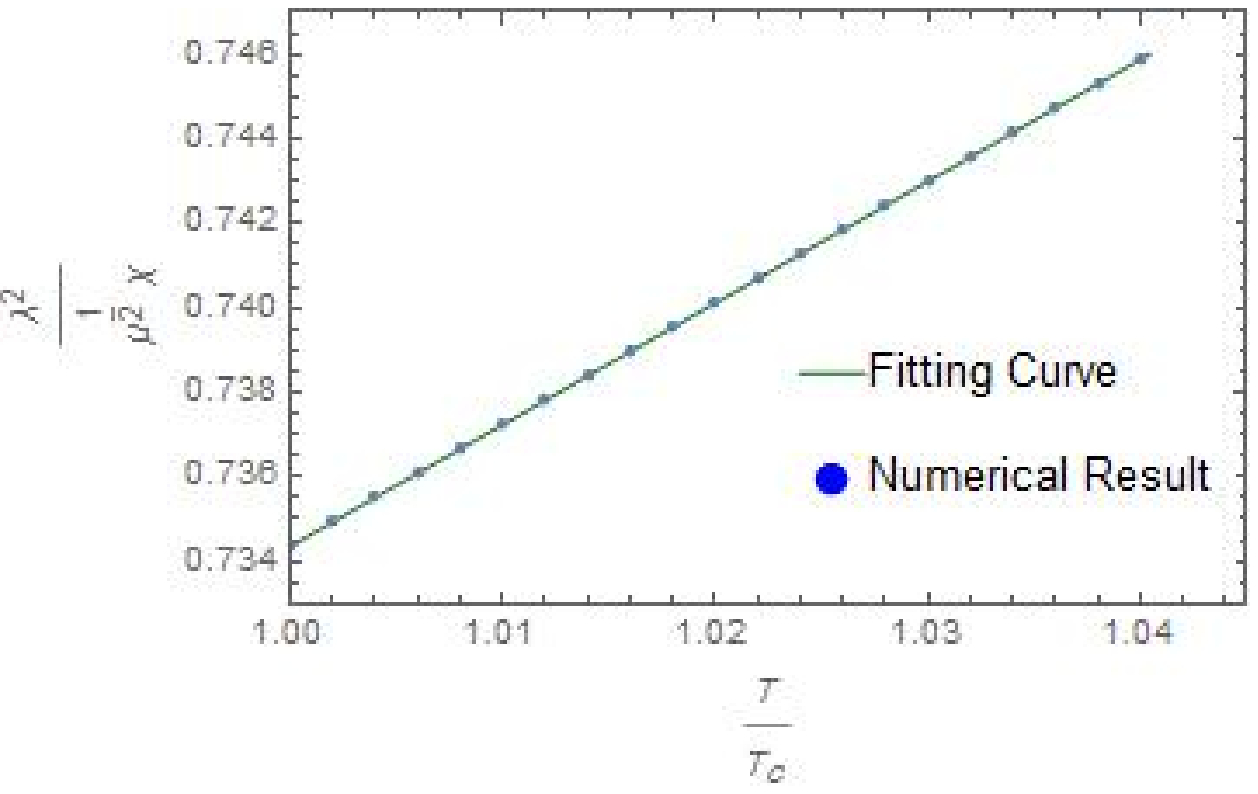}\qquad}%
            \subfigure[~$\protect$$q=11/10,\alpha=$0]{
                \label{fig32c}\includegraphics[width=.26\textwidth]{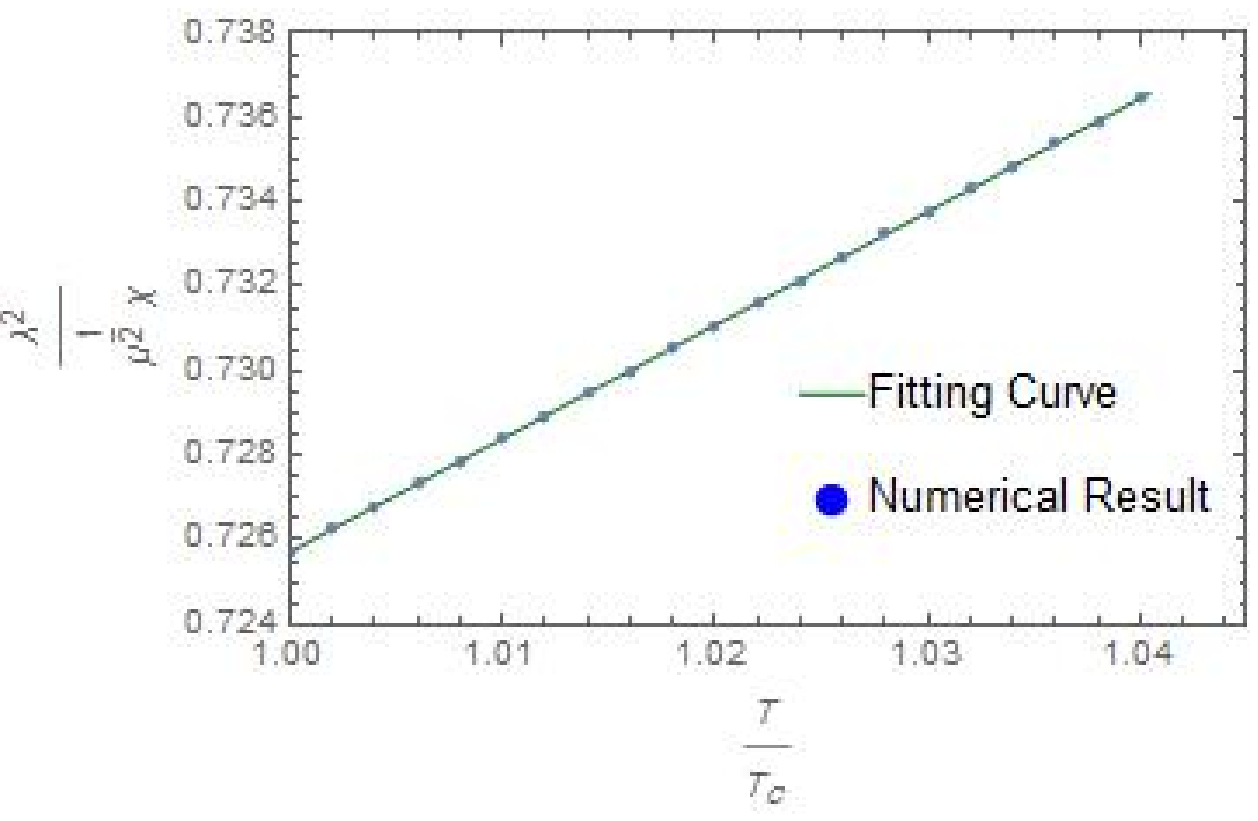}\qquad}
            \subfigure[~$\protect$$q=9/10,\alpha=$0.1]{
                \label{fig32d}\includegraphics[width=.26\textwidth]{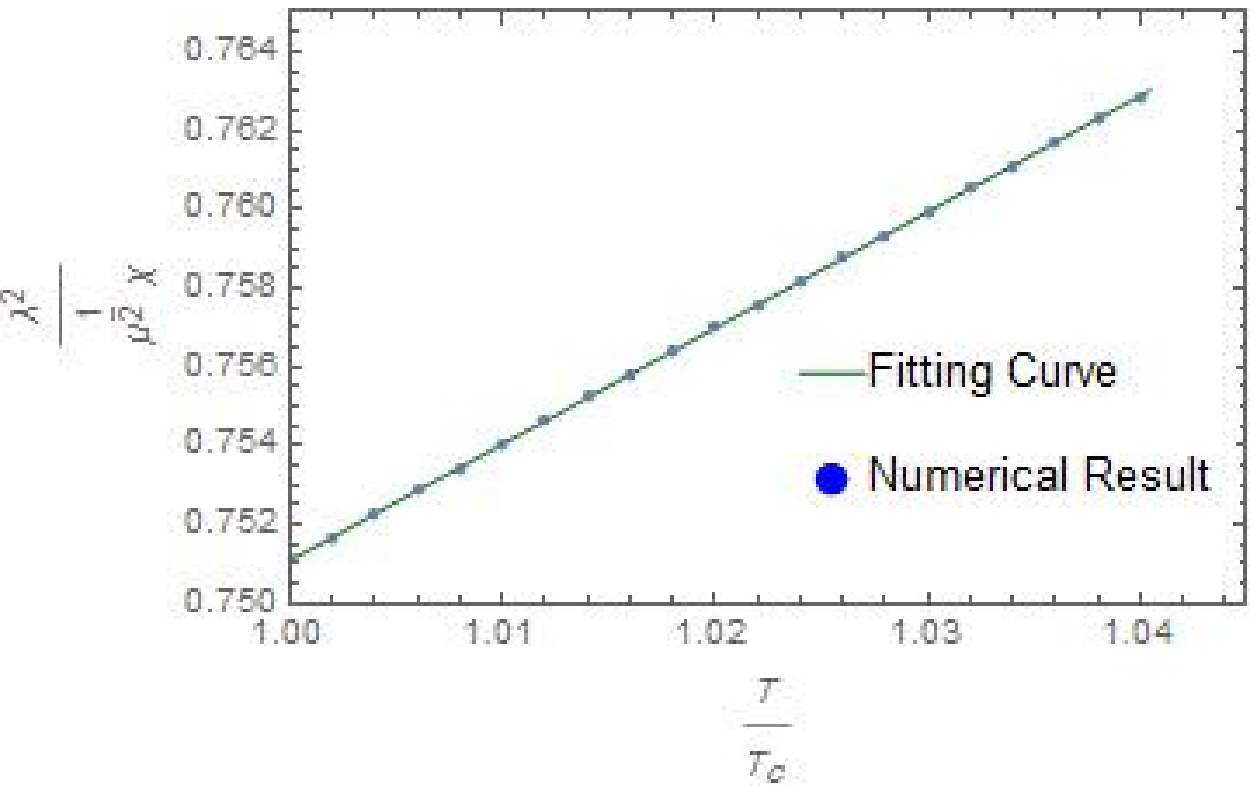}\qquad}
            \subfigure[~$\protect$$q=1,\alpha=$0.1]{
                \label{fig32e}\includegraphics[width=.26\textwidth]{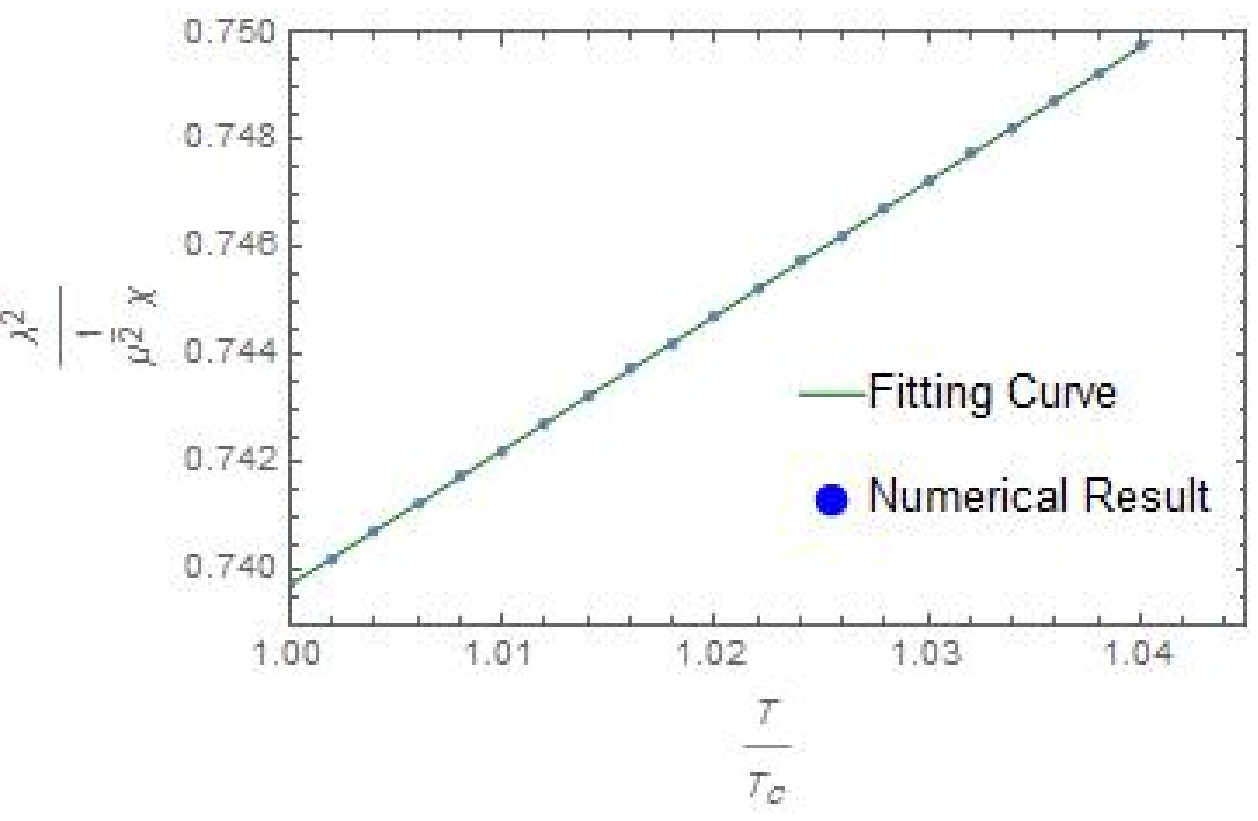}\qquad}%
            \subfigure[~$\protect$$q=11/10,\alpha=$0.1]{
                \label{fig32f}\includegraphics[width=.26\textwidth]{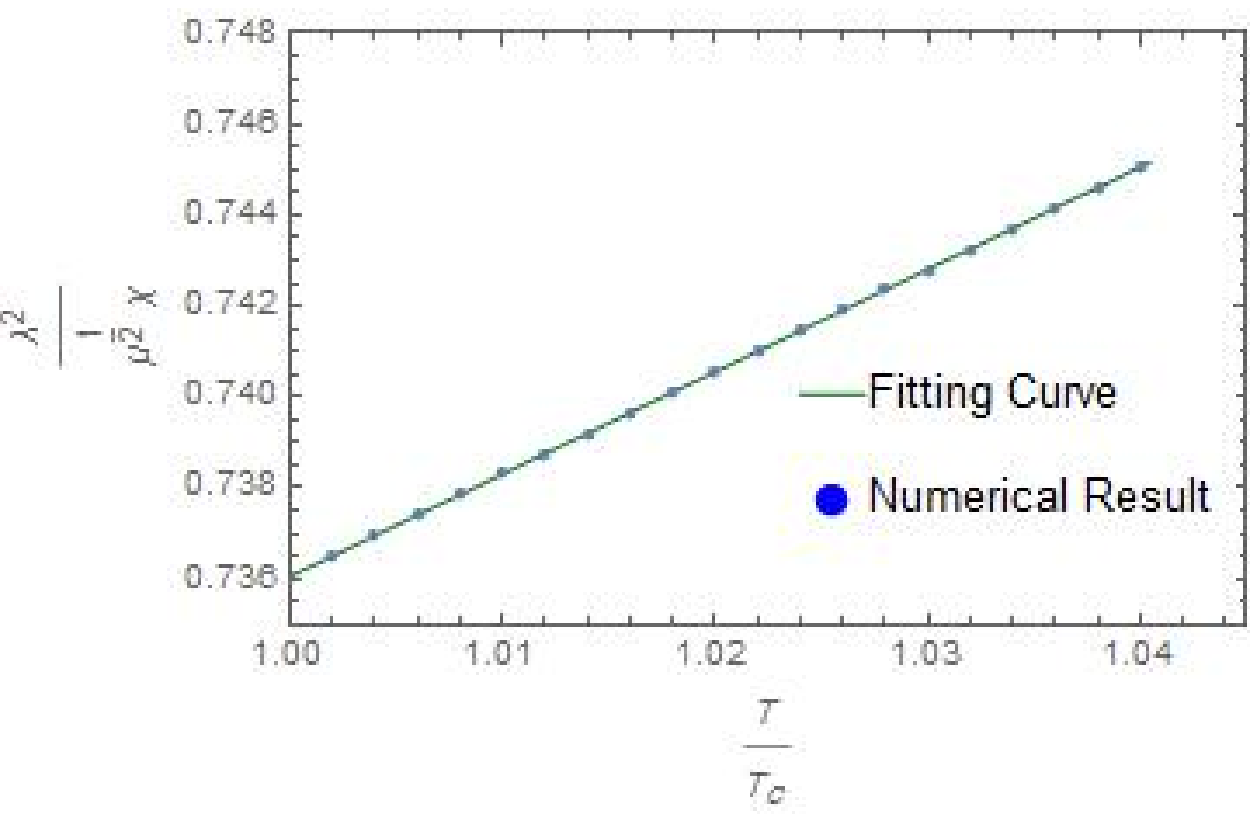}\qquad}
            \subfigure[~$\protect$$q=9/10,\alpha=$0.2]{
                \label{fig32h}\includegraphics[width=.26\textwidth]{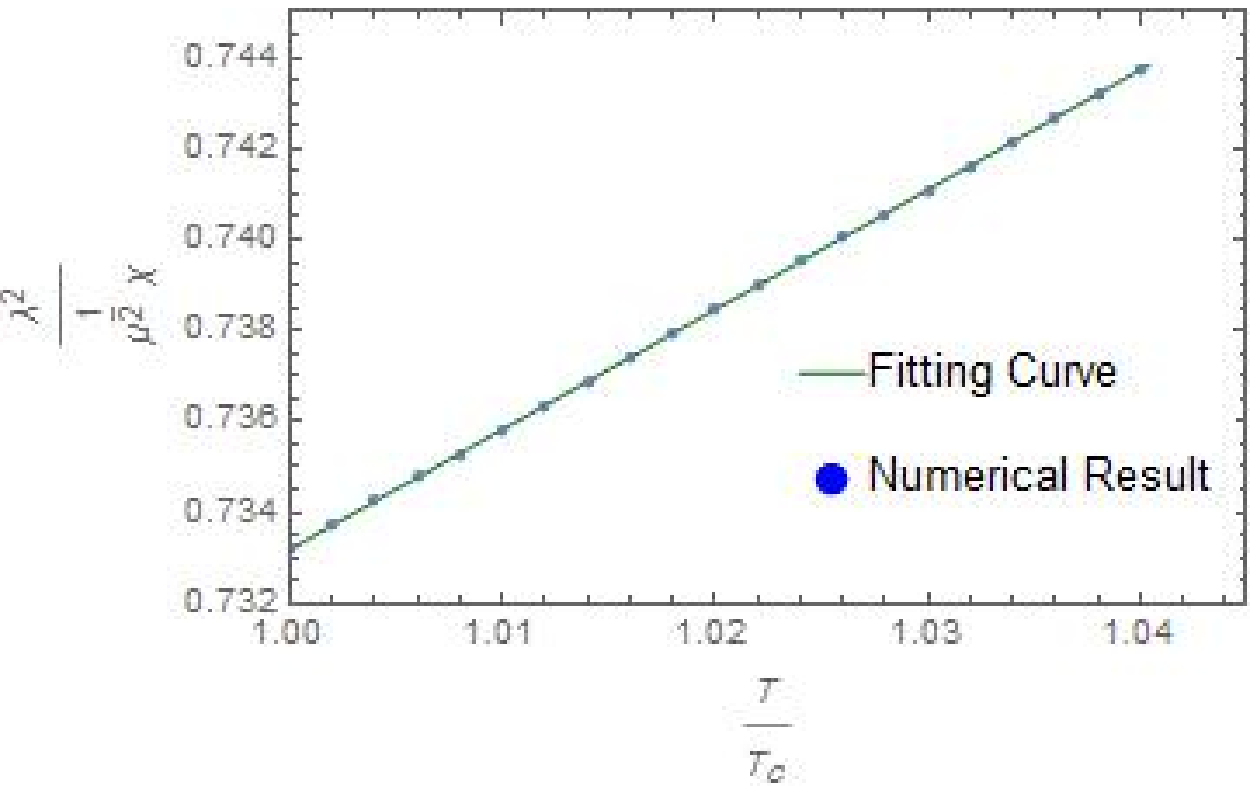}\qquad}
            \subfigure[~$\protect$$q=1,\alpha=$0.2]{
                \label{fig32t}\includegraphics[width=.26\textwidth]{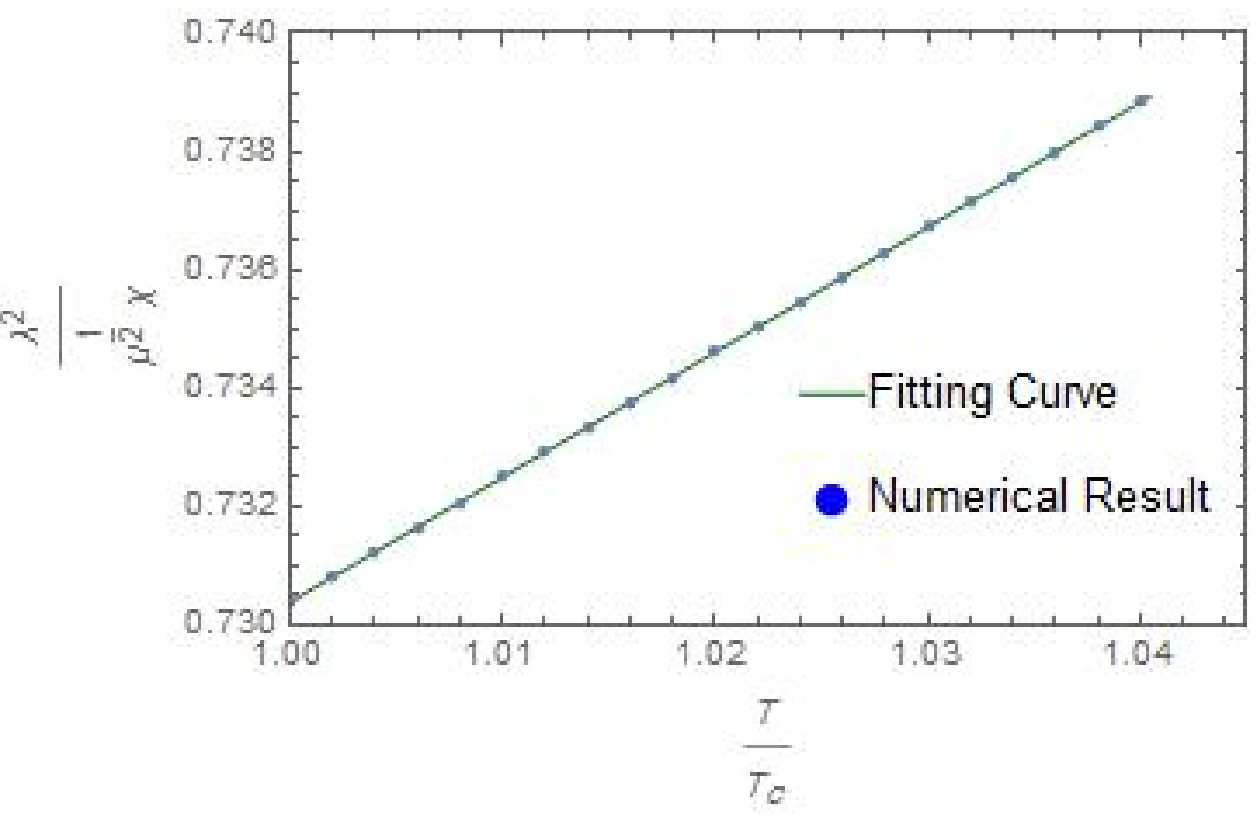}\qquad}%
            \subfigure[~$\protect$$q=11/10,\alpha=$0.2]{
                \label{fig32r}\includegraphics[width=.26\textwidth]{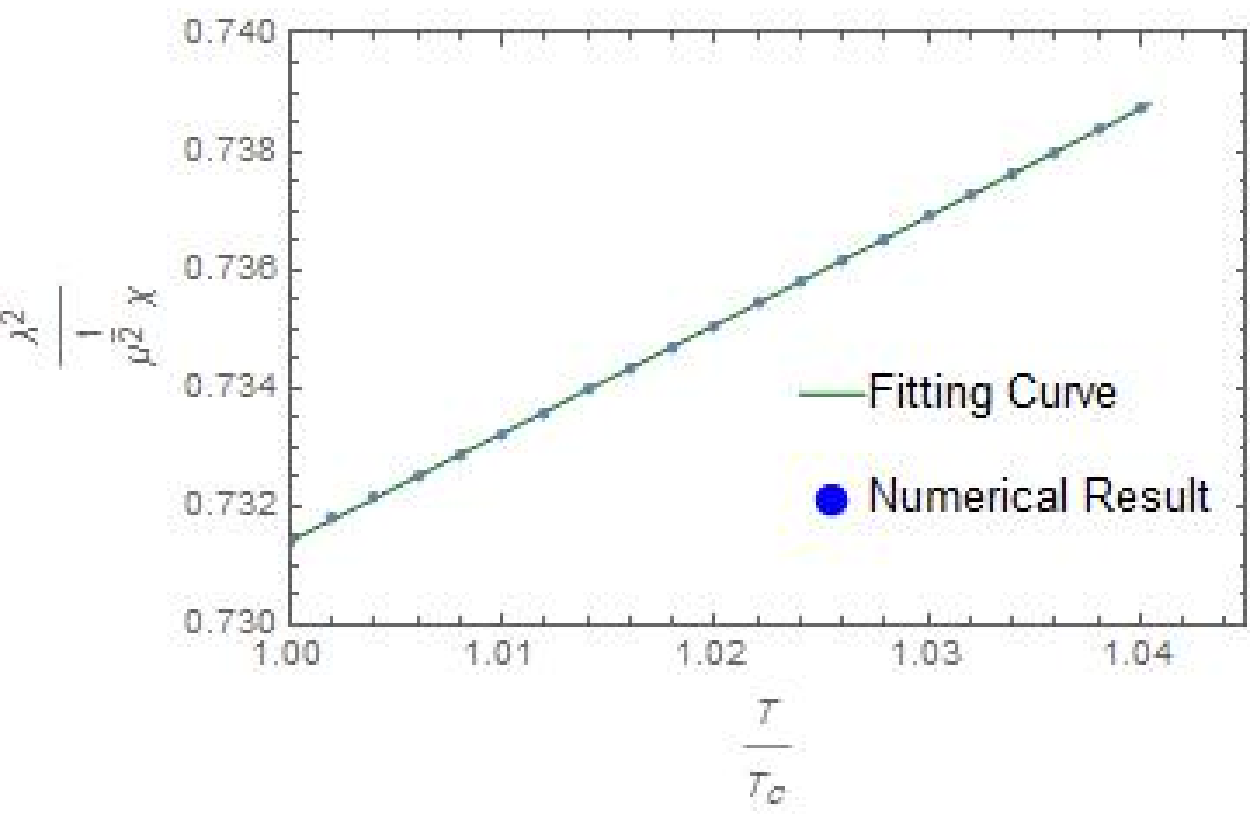}\qquad}}%

        \caption{The behavior of the inverse of susceptibility density
            $1/\chi$ with respect to the critical temperature for different
            values of $\alpha$ and $q$ in $5D$. Here we
            have taken $n=4$, $m^{2}=1/8$ and $J=-1/8$.}
        \label{fig32}
    \end{figure*}

    \section{Spontaneous magnetization \label{numst}}
    In this paper, we confined our investigation to grand canonical ensemble by considering a fixed chemical potential $\mu$. Here we employ the shooting method \cite{hartnoll} to numerically
    investigate the behavior of the holographic
    ferromagnetic-paramagnetic phase transition in $GB$ gravity. Our system has the following scaling symmetry:
     \begin{equation*}
    r\rightarrow ar,\text{ \ \ \ \ }f\rightarrow a^{2}f,\text{ \ \ \ \ }\phi
    \rightarrow a\phi,\text{ \ \ \ \ }\rho
    \rightarrow a^{2}\rho ,
    \end{equation*}%
   we can use these above scaling symmetry to obtain the solutions of Eqs. (\ref{EOM}) with the same chemical potential.
   In the follow, we choose
    $m^{2}=-J=1/8$ and $\lambda=1/2$ as a typical example in the
   numerical computation and determine the basic features of the model. Since the spontaneous magnetization in low temperature corresponds to order parameter $\rho$ in the absence of  external magnetic field, we set $B=0$ and solve Eqs.(\ref{EOM}) to get these solution for the order parameter,$\rho$, and then compute the value of the magnetic moment $N$, which is defined as
    \begin{equation}
    N=-\lambda^{2}\int\frac{\rho}{2 r^{n-1}}dr.
    \end{equation}
    Now we introduce a new variable $z=r_{+}/r$
    instead of $r$ which transforms the coordinate $r$ to dimensionless coordinate $z$. In this new coordinate, $z=0$ and $z=1$ correspond to the
    boundary ($r\rightarrow \infty$) and horizon ($r=r_{+}$),
    respectively. For convenience, we will set $l=1$ and $r_{+}=1$ in the following calculation.
    In order to investigate the trend of the magnetic moment numerically,
    we use the shooting method and solve the Eqs.(\ref{EOM}).

    We do our numerical calculation in five spacetime dimensions with $n=4$, for the cases of different $PM$ parameter $q$ and $GB$ coefficient $\alpha$.
     We present our results in Figs. \ref{fig1} when $GB$ parameter is fixed for three different values of $q$. Fig. \ref{fig2} show
     the behavior of this system for three allowed values of $GB$ parameter by fixing the power parameter $q$. These figures show the behavior of  magnetic
    moment as a function of temperature for different choices of
    nonlinearity and $GB$ parameters. For all cases, it can be
    found that as the temperature decreases, the magnetization increases and the system is in the perfect order with the maximum of
    magnetization at zero temperature. Increasing both nonlinearity or $GB$ parameter, lowers the magnetization value and the critical
    temperature. In fact, the effect of larger parameters $\alpha$ and $q$ make the magnetization harder and the ferromagnetic phase
    transition happen which is in a good agreement with previous works \cite{Zhang2,Wu1}. When the
    temperature is lower than $T_{C}$, the spontaneous magnetization
    appears in the absence of external magnetic field. For all cases, by fitting this related curve for $T<T_{c}$, we find that the phase
    transition is second-order($N\propto\sqrt{1-T/T_{C}}$). The results have been presented in
    Table \ref{Table2}. According to these
    results,  neither  $GB$ or $PM$ parameters, change the critical exponent ($1/2$), which is the value according to mean field theory. In other words,
    there is a
    holographic ferromagnetic-paramagnetic phase transition
    by considering the $PM$ electrodynamics in the $GB$ gravity similar
    to the cases of nonlinear electrodynamics in Einstein gravity
    discussed in Refs.~\cite{Zhang2, binaei1}. Table \ref{Table1}
    gives information about the values of critical temperature based
    on $\mu$ for different values of $GB$ parameter as well as power
    parameter $q$. Considering the $GB$ coefficient leads to the same
    effect as the larger values of power parameter $q$ on the critical
    temperatures. On the other hand, increasing $\alpha$ in the allowed range  can cause the transition to
        ferromagnetic phase to be harder for any values of the power
        parameter $q$.  We see also from Table \ref{Table1} that by
    increasing the power parameter $q$, the critical temperature
    $T_{c}$ decreases for fixed value of $\alpha$. It means that
    the magnetic moment is harder to be formed and the phase transition is made harder in the
    Einstein- Gauss-Bonnet gravity. This behavior have been reported
    previously in Ref. \cite{Zhang2} for nonlinear electrodynamics in
    Einstein gravity too.
  \section{ ferromagnetic material in the presence of external magnetic field \label{suscept}}
   In this section, we investigate the effect of external magnetic field for our model. So by turning on this external field,
   we examine the susceptibility density of the materials as a response to magnetic moment.
    This behavior is an important property of ferromagnetic materials.
     In order to study the static susceptibility
    density of the ferromagnetic materials in the $GB$ gravity, we follow the
    definition
    \begin{equation}
    \chi=\lim\limits_{B\to 0}\frac{\partial N}{\partial B}.
    \end{equation}
    In the presence of magnetic field, the magnetic susceptibility obtained by solving Eq.(\ref{01}). We follow the previous analysis which one has been discussed in Ref. \cite{Cai3}.
    In order to study the effect of nonlinear electrodynamics and $GB$ parameter, on the susceptibility density near the critical temperature, we plot these behaviors in Fig.\ref{fig32}. These figures show the behavior of susceptibility density as a function of the temperature for different choices of nonlinearity and Gauss–Bonnet parameters.
    We see that increasing each one of these parameters, makes the susceptibility value increases when the temperature decreases. In fact, with increasing these parameters,
     the system will become unstable and then the ferromagnetic phase will be broken into the paramagnetic phase. In the
    region of $T\to T^{+}_{c}$, the susceptibility density satisfies
    the Curie-Weiss law of ferromagnetic and the paramagnetic phase
    happens,
    \begin{equation}
    \chi=\frac{C}{T+\theta},\text{ \ \ \ \ } T>T_{C},
    \end{equation}
    where $C$ and $\theta$ are two constants. The numerical
    results are listed in Table \ref{table3}. Obviously, we can see that the
    coefficient in front of $T/T_{c}$ for $1/\chi$ increases, by
    decreasing the power parameter($q$) and $GB$ coefficient($\alpha$).
    It means that for the smaller values of these two parameters
    our system becomes stable.

    \begin{table*}[ht]
        \centering%
        \caption{The magnetic susceptibility $\chi$ with different values
            of $q$ and $\alpha$.}
        \begin{tabular}{llllll}
            \hline
            & $q$  & $0.9 $ & $1$ & $11/10$ \\
            \hline
            $\alpha=0$&  $\lambda^{2}/{{\chi}{\mu}}$ & $0.3236(T/Tc-1.331)$  & $0.2881(T/Tc-1.5486)$  & $0.2683(T/Tc-1.7039)$  \\
            & ${\theta/\mu}$  & $3.5675$  & $3.7735$ & $3.9491$  \\
            $\alpha=0.1$& $\lambda^{2}/{{\chi}{\mu}}$   & $0.2947(T/Tc-1.5483)$  & $0.2500(T/Tc-1.9584)$& $0.2251(T/Tc-2.2696)$  \\
            & ${\theta/\mu}$  & $3.9253$  & $4.5182$ & $4.9790$ \\
            $\alpha=0.2$& $\lambda^{2}/{{\chi}{\mu}}$  & $0.2635(T/Tc-1.7828)$  & $0.2114(T/Tc-2.4542)$& $0.1828(T/Tc-3.0002)$  \\
            & ${\theta/\mu}$ & $4.1108$  & $5.1496$ & $5.9833$ \\
            \hline
        \end{tabular}
        \label{table3}

    \end{table*}


    \section{Closing remarks}
    We have numerically investigated the behavior of a holographic
    ferromagnetic model with the $PM$ electrodynamics based on
    shooting method, by considering the higher order \emph{GB}
    correction terms on the gravity side of the action. On the gauge
    field side, however, we have considered the effects of \emph{PM}
    nonlinear electrodynamics on the system. We have focused on
    $1/2<q<n/2$ as the physical range of the power parameter. We found
    that for this system, increasing the value of nonlinearity
    parameter and/or \emph{GB} coefficient lowers the critical temperature. Numerical calculations indicate that
    increasing the values of nonlinearity and \emph{GB} parameters in
    different dimensions can make magnetization harder to be
    formed, since the increasing $\alpha$ and $q$ always inhibit the ferromagnetic phase transition.
    Increasing the effect of \emph{PM} parameter in \emph{GB} gravity
    leads to the same behavior as in case of Einstein gravity
    \cite{binaei1}. We observed that the enhancement in \emph{GB}
    parameter $\alpha$ causes the paramagnetic phase more difficult to
    appear. These results are reflected in
    Figs.\ref{fig2}. We find out
    that the magnetic moment behaves as $ (1-T/T_{c})^{1/2}$ indicating that the critical exponent has the mean field value ($1/2$), which seems to be a
    universal constant, as it it the same value for the Einstein case. We conclude that this critical exponent is not effected by
    the model parameters such as $q$, $\alpha$ and $n$.

    In the presence of external magnetic field, the inverse magnetic
    susceptibility near the critical point behaves as
    ($\frac{C}{T+\theta}$) for all allowed values of the power
    parameter $q$ and different values of the \emph{GB} coupling $\alpha$ in
    different dimensions, and therefore it satisfies the Curie-Weiss
    law. The absolute value of $\theta$ increases by increasing each one of $q$ and $\alpha$. When $T<T_{c}$ the ferromagnetic phase happens, and for
    $T>T_{c}$ this model goes to the paramagnetic phase. As a result,
    our model provides a holographic description for the
    ferromagnetic-paramagnetic phase transition.
    \begin{acknowledgments}
        {We thank the Research Council of Shiraz University.}
    \end{acknowledgments}

\end{document}